\documentclass[12pt,draft]{article}
\usepackage{psfig}
\usepackage[usenames]{color}

\makeatletter
\@addtoreset{equation}{section}

\makeatletter
\renewcommand\section{\@startsection {section}{1}{\z@}%
                                   {-3.5ex \@plus -1ex \@minus -.2ex}
                                   {2.3ex \@plus.2ex}%
                                   {\normalfont\large\bfseries}}
\renewcommand\subsection{\@startsection{subsection}{2}{\z@}%
                                     {-3.25ex\@plus -1ex \@minus -.2ex}%
                                     {1.5ex \@plus .2ex}%
                                     {\normalfont\bfseries}}

\def\baselinestretch{1.2}
\newcommand{\be}{\begin{equation}}
\newcommand{\ee}{\end{equation}}
\newcommand{\beq}{\begin{eqnarray}}
\newcommand{\eeq}{\end{eqnarray}}

\def\[{\left [}
\def\]{\right ]}
\def\({\left (}
\def\){\right )}
\def\p{\partial}
\def\R{{\bf R}}

\def\Z{{\bf Z}}

\def\p{\partial}

\def\r2{\sqrt{2}}
\def\M{{\cal M}}
\def\A{{\cal A}}
\def\p{{(p)}}
\def\pp{{(pp')}}

\begin{document}
\begin{titlepage}

\begin{flushright}
{\small UNH-05-03, UCB-PTH-05/16\\
UPR-1119-T, NSF-KITP-05-29 \\ 
hep-th/0505167}
\end{flushright}


\vfil\

\begin{center}

{\Large{\bf Supergravity Microstates for BPS Black Holes and Black~Rings}}
\vfil

\vspace{3mm}

Per Berglund\footnote{e-mail:per.berglund@unh.edu}$^{,a}$, Eric G.
Gimon\footnote{e-mail:eggimon@lbl.gov}$^{,b}$$^{,c}$ and Thomas S.
Levi\footnote{e-mail:tslevi@sas.upenn.edu}$^{,d,e}$
\\

\vspace{8mm}

\bigskip\medskip\centerline{$^a$ \it Department of Physics, University
 of New Hampshire,
 Durham, NH 03824, USA.} \smallskip \centerline{$^b$
\it Department of Physics, University of California,
Berkeley, CA 94720, USA.}
 \smallskip\centerline{$^c$
\it Theoretical Physics Group, LBNL, Berkeley, CA 94720, USA.}
\smallskip \centerline{$^d$ \it David Rittenhouse Laboratories,
  University of Pennsylvania, Philadelphia, PA 19104, USA.}
\smallskip \centerline{$^e$ \it Kavli Institute for Theoretical
  Physics,
University of California, Santa Barbara, CA, 93106, USA.}

\vfil

\end{center}
\setcounter{footnote}{0}
\begin{abstract}
\noindent
We demonstrate a solution generating technique, modulo some
constraints, for a large class of smooth supergravity solutions with
the same asymptotic charges as a five dimensional 3-charge BPS black
hole or black ring, dual to a $D1/D5/P$ system. These solutions are
characterized by a harmonic function with both positive and negative
poles, which induces a geometric transition whereby singular sources
have disappeared and all of the net charge at infinity is sourced by
fluxes through two-cycles joining the poles of the harmonic
function.
\end{abstract}
\vspace{0.5in}

\end{titlepage}
\renewcommand{\baselinestretch}{1.05}  
\section{Introduction and Summary}

The black hole information puzzle \cite{hawking} is a particularly
striking example of the problems encountered when trying to combine
quantum mechanics and gravity. Various computations in string theory
have suggested that the usual picture of a black hole with an event
horizon, then empty space with a central singularity, could be an
emergent phenomena arising when we coarse grain over a set of
microstates. A similar conjecture applies for the new black ring
solutions \cite{henring, henring1, matring1, matring2, EEMR, reall1,
emparan1}. Recently, Mathur and Lunin have made a stronger
conjecture~\cite{mathurmultstring}\ (see \cite{mathurreview} for a
recent review) which holds that black hole microstates are
characterized by string theory backgrounds with no horizons.

Up to now the main evidence for this conjecture involves finding
microstates for two charge proto-black holes with no classical
horizon area.  Recently, Mathur et al \cite{mathur1,mathur2,mathur3}
uncovered some three charge supergravity solutions with no horizon (see also \cite{oleg}).
Since these solutions exactly saturate a bound on angular momentum,
they have the same asymptotic charges as a black hole whose horizon
area classically vanishes. In \cite{rossnonsusy} some finite
temperature (non-supersymmetric) generalizations appear for these
which correspond to black holes with finite size horizons; we are
interested in finding microstates for BPS black holes which have
already have finite size horizons for zero temperature.

   Bena and Warner \cite{BW} (see also \cite{GR}) have developed a
formalism for finding five dimensional supersymmetric solutions with
three asymptotic charges and two angular momenta as reductions from
M-theory. We will show how a general class of these solutions can be
layed out which corresponds (modulo some consistency conditions) to
$U(1)$ invariant microstates with the asymptotics of black objects
with finite areas, dual to the $D1/D5/P$ system. We develop the
simplest cases; some too simple, in fact, to look like black objects
with non-vanishing horizons.  The framework we lay out, however,
exhibits strong potential for finding $U(1)$-invariant supergravity
microstates for all five-dimensional black objects.

   The Bena-Warner ansatz involves a fibration of the time coordinate
over a hyperkahler base space.  The key insight from \cite{mathur3}
is that even if this base space is singular, a time fibration can be
formulated such that the total space is completely smooth.  This
gives us access to a whole new class of base spaces: new two-cycles
replace the origin of $\R^4$.

   We will review the Bena-Warner ansatz in section \ref{eqs} and then
in section~\ref{sols} we will offer a general form for a solution to
these equations (a similar set of ansatz appears in
\cite{denef1,denef2,denef3,GSY, EEMR, BKW}) which are obviously
smooth everywhere except at some orbifold points and possibly where
the base space is singular.  In section~\ref{smooth} we will
demonstrate that our solution is smooth everywhere with no closed
timelike curves or event horizons given three general sets of
constraints. In section \ref{basic} we outline a simple class of
examples for our formalism related to the solutions in
\cite{mathur1,mathur2,mathur3}.  They saturate a bound on rotation,
and so correspond to microstates of black holes with no classical
horizon.  In section~\ref{more} we discuss how these examples can be
generalized so as to include microstates for black holes which have
finite size horizons.  Finally we will briefly discuss further
generalizations and future directions.

Work concurrent with this paper is also appearing in~\cite{BW2}.

\section{The Bena-Warner Ansatz and System of Equations}
\label{eqs}

     In \cite{BW}
Bena and Warner neatly lay out an ansatz for a 1/8 BPS solution with
three charges in five dimensions as a reduction from M-theory where
the charges come from wrapping membranes on three separate $T^2$'s
used to reduce from eleven dimensions.  The five dimensional space
is written as time fibered over a hyperkahler base space, $H\!K$.
The resulting M-theory metric takes the form:
\beq
\label{m-theory metric}
ds^2 _{11}= -(Z_1 Z_2 Z_3)^{-2/3}( dt + k)^2+(Z_1 Z_2 Z_3)^{1/3} ds^2_{HK}
 + ds^2 _{T^6} ,
\eeq
where
\beq
ds^2_{T^6} &=& (Z_1Z_2 Z_3)^{1/3} \(Z_1^{-1} (dz_1 ^2 +dz_2^2) +
Z_2^{-1} (dz_3 ^2 +dz_4^2) + Z_3 ^{-1} (dz_5 ^2 +dz_6^2)\).
\eeq
The $Z_i$'s and $k$ are functions and one-form respectively on the
hyperkahler base space, the three $T^2$'s have volumes $V_i$.  The
gauge field takes the form:
\beq
\label{cfield}
 C_{(3)} &=& - (dt + k) \( Z_1^{-1}\, dz_1\wedge dz_2 +
Z_2^{-1}\, dz_3\wedge dz_4 + Z_3^{-1}\,
dz_5\wedge dz_6 \) \nonumber \\
&& + 2 \,a_1 \wedge  dz_1\wedge dz_2 + 2\, a_2 \wedge  dz_3\wedge
dz_4 + 2\, a_3 \wedge  dz_5\wedge dz_6 ,
\eeq
where the $a_i$ are one-forms on the base space.  After reduction on
$T^6$ the C-field effectively defines three separate $U(1)$ bundles,
with connections ${\cal A}_i = -(dt + k)Z^{-1}_i + 2\,a_i$, on the
5-dimensional total space. If we now define the two-forms,
\beq
G_i = da_i,
\eeq
\cite{BW} show that the equations of motion reduce to the three conditions
(here the Hodge operator refers only to the base space $H\!K$):
\beq
\label{eq1} G_i &=& \star  G_i, \\
\label{eq2} d\!\star \!dZ_i &=& 2s^{ijk}\, G_j \wedge G_k, \\
\label{eq3} dk + \star dk &=& 2\, G_iZ_i.
\eeq
where we define the symmetric tensor $s^{ijk} = |\epsilon^{ijk}|$.

\section{Solving the Equations and Asymptotics}
\label{sols}

    In this section, we will first solve the Bena-Warner ansatz by
selecting a base space $H\!K$ which has the property that the
reduced five dimensional space is asymptotically flat.  Written in
Gibbons-Hawking form, the metric for this base space is:

\be
\label{HK}
ds^2_{HK} = H^{-1} \sigma^2 + H (dr^2 + r^2 d\theta^2 +
r^2 \sin^2\theta
 d\phi^2),
\ee
where $H$ is a positive function on $\R^3$ with integer poles and
$\sigma$ is a one form on $\R^3$ of the form $d\tau + f_adx^a$ ($\tau$ has period
$4\pi$) satisfying
\be
\star _3 d\sigma = dH.
\ee
Asymptotic flatness with no $ADE$ identification basically forces
the choice $H = 1/r$ on us, this is fairly limiting.  Following
\cite{mathur3}, however, we know that the time fibration can allow us to
relax the hyperkahler condition on $H\!K$.  We allow a singular
pseudo-hyperkahler $H\!K$ (see \cite{mathur3} for a definition) so long as the total space with the
time-fiber is smooth.

Our complete solution will be encoded using a set of 8 harmonic
function on $\R^3$: $H, h_i, M_i$ and $K$, generalizing the
appendix in \cite{mathur3} patterned on \cite{allsols}.  These can take a fairly
generic form, with constraints and relations that we will gradually
lay out and then summarize.

\subsection{The function $H$}

Relaxing the hyperkahler conditions provides us with more general
candidates for $H$, such as:
\be
\label{H}
H = \sum_{p\,=1}^{N} \frac{n_p}{r_p}, \qquad r_p =
|\vec{r}_p| = |\vec{x} - \vec{x}_p|, \qquad \sum_{p\,=1}^{N} n_p =
1.
\ee
The last sum guarantees that asymptotically we have an $\R^4$.
Relaxing the condition that the $n_p$ be positive has allowed us to
introduce an arbitrary number of poles, $N$, at positions
$\vec{x}_p$. We can use this potential to divide $\R_3$ into a part
where $H>0$ (Region I) and a compact (but not necessarily connected)
part where $H<0$ (Region II). The two are separated by a domain wall
where $H=0$ (from now we will refer to this as ``the domain wall'').
The metric in eq.({\ref{HK}) is singular at this domain wall, but we
will see in the next section how the $t$-fibration saves the day. It
does not, however, change the fact that with our choice of $H$ there
are two-cycles, $S^{pq}$, coming from the fiber $\sigma$ over each
interval from $\vec{x_p}$ to $\vec{x_q}$; it just makes them
non-singular.

For convenience we define the following quantities:
\be
\label{ps}
\Pi^r = \prod_{p\,=1}^N r_p,\;\; \Pi^r_{p} = \prod_{q\ne p}
r_{q},\;\; \Pi^r_{ps} = \prod_{s \ne q \ne p}\!\!r_q, \qquad f =
\sum_{p\,=1}^{N} n_p \Pi^r_p.
\ee
Hence, $H$ can be written as $f/\Pi^r$.  Note also that eq.(\ref{H})
implies  that $f \to r^{N-1}$ at asymptotic infinity, regardless of
our choice of origin.

\subsection{The Dipole Fields}
To generate the $G_i$, it is useful to define harmonic functions on
$H\!K$ that, up to a gauge transformation, fall off faster than $1/r$ as $r\to\infty$.  To limit the
scope of our discussion, we choose these to have poles in the same
place as $H$\footnote{It is our sense that poles in other places
lead to superpositions with previously known solutions such as
supertubes and AdS throats and so would distract attention from the
new solutions we present}.
\be \label{d}
h_i = \sum_{p\,=1}^{N} \frac{d_i^{(p)}}{4{r}_p}\,, \quad
\sum_{p\,=1}^{N} d_i^{(p)} =0.
\ee
With these functions we define self-dual $G_i$'s:
\be \label{gauge}
 G_i = d(h_i/H)\wedge \sigma - H\!\star _3d(h_i/H),
\ee
which have the requisite fall-off ($\star_3$ is the Hodge operator
on $\R^3$). Note that we can always satisfy the condition in eq.(\ref{d}) by applying the gauge transformation
\be
h_i \to h_i - {1 \over 4} \sum_p d^{(p)} _i \, H .
\ee
We can integrate the two-forms partially to get an expression
for the $a_i$'s:
\beq
  a_i = (h_i/H) \,\sigma + a_{ia} dx^a, \qquad d(a_{ia} dx^a) =
  -\star _3dh_i
\eeq
where the $dx^a$ denote any complete set of one-forms on flat
$\R^3$. Since we have chosen to localize the poles of the $h_i$ on
top of the poles of $H$, these one-forms have no singularities
except at $H=0$.  It useful to define the dipole moments:
\be
\label{dipoles}
 \vec{D}_i = \sum_{p\,=1}^{N} d^\p_i \vec{x}_p,
\ee
which control the asymptotics of our solution.  Note that a priori
these need not be parallel vectors such as in the asymptotics of a
single black ring.  Also, since $\sum_p d_i^\p =0$, these dipole
moments are independent of the choice of origin for $\R^3$. It is
also useful to define, more locally, the relative dipole moments and
to rewrite expressions as a function of these quantities:
\be
d_i^{pq} = n_pd_i^{(q)} -  n_qd_i^{(p)}\;\;\; \Rightarrow\qquad
d_i^{(p)} = -\sum_{q=1}^N d_i^{pq}, \qquad \vec{D}_i = - {1\over 2}
\sum_{pq} d_i^{pq}\,(\vec{x}_p - \vec{x}_q).
\ee
The $d_i^{pq}$'s will appear almost everywhere in our solution.  As
we will see, they measure the various $U(1)$ fluxes through the
two-cycle, $S^{pq}$, connecting $\vec{x}_p$ and $\vec{x}_q$.

\subsection{The Monopole fields}

The membrane charge at infinity can be read off from the C-field
components with a time component.  Looking at eq.(\ref{cfield}) this
means that the $Z_i$'s encode the three membrane charges $Q_i$,
therefore they must have a falloff like
\be
Z_i \to 1 + {Q_i\over 4r}\qquad \textrm{as}\qquad r \to \infty.
\ee
since the asymptotic $\R^4$ radial coordinate is $R = 2r^{1\over 2}$.
Taking advantage of the natural radial distances in $H$ we can
write arbitrary new harmonic functions,
\be
M_i = 1 + \sum_{p\,=1}^{N} {Q^\p_i\over 4r_p},\qquad
\sum_{p\,=1}^{N} Q_i^\p = Q_i,
\ee
which have exactly the asymptotics above.  We use these to write down
the following ansatz for the $Z_i$'s
\beq
Z_i &=& M_i + 2\, H^{-1}s^{ijk}h_j\,h_k.
\eeq
The correction terms fall off like $1/r^2$ so they don't mess up the
asymptotics.  They are necessary to satisfy the second equation of
motion, eq.(\ref{eq2}):
\be
\label{eom2}
d\!\star \!dZ_1 = 4\, G_2\wedge G_3,\qquad  d\!\star \!d Z_2 = 4\,
G_1\wedge G_3,\qquad d\!\star \!dZ_3 = 4\, G_1\wedge G_2 ,
\ee
up to some extra delta function sources.  For the purposes of this
paper, we will focus our attention on $Z_i$'s tuned so that they
have no singularities apart from those at $H=0$.  This removes the
delta function sources in eq.(\ref{eom2}) and allows us to write the
$Q_i^\p$ explicitly as:
\be
\label{fixq}
Q_i^\p = - {s^{ijk}\over 2n_p}\,{d_j^\p\,d_k^\p}.
\ee
We can now rewrite the $Z_i$ in the following form:
\be
\label{zform} Z_i = 1 - {s^{ijk}\over 4f} \sum_{p,q}
\frac{d_j^{pq}d_k^{pq} \Pi^r_{pq}}{4n_pn_{q}},
\ee
where we have used the quantities $f$ and $\Pi^r_p$ from
eq.(\ref{ps}). This implies an alternate form for the $Q_i$,
\be
Q_i = \sum_{p,q} Q_i^{pq}, \qquad {\rm where}\qquad Q_i^{pq} = 
-
 \frac{s^{ijk} d_j^{pq}d_k^{pq}} {4n_pn_q},
\ee
which makes clear that we can also think of the total charge at
infinity as coming from a sum of contributions from each two-cycle
$S^{pq}$.

\subsection{The Angular Momentum}

Finally, lets look at the angular momentum.  We use the natural
basis
\be
k = k_0 \,\sigma + k_a\, dx^a,
\ee
with the $d\alpha^i$ a complete basis of one-forms on $\R^3$. Our
ansatz is now:
\beq
\label{k0}
k_0 &=& K + 8 H^{-2}\,h_1\,h_2\,h_3 + H^{-1}\,M_1\,h_1
+ H^{-1}\,M_2\,h_2 +  H^{-1}\,M_3\,h_3  \\
    &=& K - 4 H^{-2}\,h_1\,h_2\,h_3 + Z_i\,(h_i/H) = \bar{K} + Z_i\,(h_i/H)\nonumber\\
\label{star}
\star _3 d(k_adx^a) &=& H dK - K\,dH + h_i\, dM_i - M_i\,dh_i \\
      &=& H\, d\bar{K} - \bar{K}\,dH + h_i\,dZ_i - Z_i\,dh_i
\nonumber
\eeq
Here we have defined a new harmonic function, $K$, and its partner
function, $\bar{K}$:\footnote{$\ell_p$ should not be confused with
the Planck length $\ell_P$.}
\be
K = \sum_{p\,=1}^{N}\left({\ell_p\over r_p}\right),\qquad
 \bar{K} = K - 4 H^{-2}\,h_1\,h_2\,h_3.
\ee
The regularity of $k$ is important here.  There is an integrability
condition on eq.~(\ref{star}) which basically requires that
``$d\star _3$'' of that equation is zero; this is trivially
satisfied everywhere except at the poles of our harmonic functions.
We would like the one-form $k$ to have no singularities except at
$H=0$ and for $k_adx^a$ to be a globally well defined one-form
everywhere on $\R^3$ with an asymptotic fall-off like $1/r$. This
turns out to be possible if we use the freedom to add any closed
form of our choice to $k_adx^a$, and demand:
\be
\label{Kpoles}
\ell_p = \frac{d^\p_1d^\p_2d^\p_3}{16n_p^2},\qquad k_0|_{r_p=0} = 0.
\ee
The first condition removes poles in $\bar{K}$, the second condition
insures that $d^2(k_adx^a=0)$ everywhere, i.e. that $k_adx^a$ is
globally well defined.  With the first condition we can rationalize
our form for $k_0$ a little bit:
\beq
k_0 &=& \sum_{p\,=1}^{N}\sum_i \frac{d_i^\p\Pi^r_p}{4 f}  + {1\over
16\Pi^r f^2}\Bigg[\sum_{p,q,s} \Big( d_1^\p d_2^\p d_3^\p
\frac{n_qn_s}{n_p^2} - d_1^\p
d_2^{(q)}d_3^{(s)}\Big)\Pi^r_p\Pi^r_q\Pi^r_s \\
&&- s^{ijk}\sum_{p\,=1}^{N} d_i^\p \Pi^r \Pi^r_p \sum_{q,s}
\frac{d_j^{qs}d_k^{qs} \Pi^r_{qs}}{4n_qn_{s}} \Bigg] \nonumber
\eeq
This allows us to solve for $k_0=0$ at the points ${r}_p =0 $, boiling down to
\be
\label{fixdistance}
0 =  \sum_i d^\p_i \; + \sum_{q} {1\over 4
n_p^2 n_q^2}{1\over r_{pq}} \prod_i
d_i^{pq}
\ee
where $r_{pq} = |\vec{x}_p - \vec{x}_q|$. This puts at most $N-1$
independent constraints on the relative positions of the poles. Of
course, all the $r_{pq}$ have to be non-negative, therefore a bad
choice of $d_i^{pq}$ may lead to no solution at all! Note, also,
that if any of the $d_i^{pq}$ vanishes, then the corresponding
$r_{pq}$ will not appear in these constraints.

 Using
eqs.(\ref{Kpoles}) and (\ref{fixdistance}), we can rewrite
eq.(\ref{star}) as:
\beq
\label{simplestar}
\star _3 d(k_adx^a) =  -\sum_{p,q} {1\over 32n_p^2n_q^2} \prod_i
{d_i^{pq} \over r_{pq}r_q^2r_p^2}\Big((r_{pq} - r_p)r_pdr_q - (r_{pq}-r_q)r_qdr_p\Big)
\eeq
If we define $\phi_{pq}$ as the right-handed angle about the directed
line from $\vec{x_p}$ to $\vec{x_q}$, we can integrate the expression above
explicitly to get:
\be
\label{kvec}
k_adx^a = \sum_{p,q} {1\over 64n_p^2n_q^2} \prod_i {d_i^{pq} \over
  r_{pq}^2 r_qr_p} (r_p + r_q - r_{pq})\Big(r_{pq}^2 - (r_p -
r_q)^2\Big)\, d\phi_{pq}.
\ee

  The two-form $dk$ naturally splits into a self-dual and
anti-self-dual part.   The split gives:
\beq
dk_L = (dk + \star \,dk)/2 &=& \,Z_iG_i \\
dk_R = (dk - \star \,dk)/2 &=& (d\bar{K}\wedge\sigma + H\!\star _3\!d\bar{K})
 + (h_i/H)\,(dZ_i\wedge\sigma + H\!\star _3\!dZ_i)
\eeq
Notice that we have correctly solved eq.(\ref{eq3}) and also that $K$ contributes
only to the anti-self-dual part of $dk$. Looking at the asymptotics,
these tensors look like:
\beq
 dk_L &\to& \sum_i G_i \to {1\over 4}\left[ d\({\sum_i4\vec{D}_i\cdot\hat{r}\over
      4r}\)\wedge\sigma - {1\over 4r}\!\star_3\!d\({\sum_i4\vec{D}_i\cdot\hat{r}\over
      r}\)\right] \\
 dk_R &\to& {\sum_{p=1}^N 16\ell_p}\,\cdot \,{1\over 4} \Big[d\({1\over 4r}\)\wedge\sigma +
      {d\sigma\over 4r}\Big] \nonumber
\eeq
{}From these expressions we can read the magnitudes of the angular
momenta as measured at infinity.  They take the values :
\beq
\label{J}
J_L &=& {4 G_{5}\over \pi } j_L = |4 \sum_i \vec{D}_i|, \\
J_R &=& {4 G_{5}\over \pi } j_R = 16\,\sum_{p=1}^N \ell_p,
\eeq
The $j_{L,R} \in \Z$ and $G_5$ is the five-dimensional Newton's
constant, which has length dimensions three and takes the value $G_5
= 16\pi^7\ell_P^9\prod_iV_i^{-1}.$

\section{Smoothness of the Solution}
\label{smooth}

So far, we have proposed general forms for the $G_i$, $Z_i$ and $k$
which satisfy the equations (\ref{eq1}-\ref{eq3}).  The next step is
to demonstrate that we can exhibit solutions without any
singularities.  Since our base space $H\!K$ is singular when $H\to
0$ it seems natural to check that the full metric and the C-field
are free of singularities near this domain wall. In this section, we will
demonstrate how, with the most general harmonic functions, we avoid
any singularities at the domain wall.  We will also
address other potential pitfalls which might render our solution unphysical.

  Assuming that none of the poles in the
$h_i,M_i$ and $K$ overlap with the domain wall, we see that $a_i,Z_i$ and
$k$ have the following expansions as $H\to 0$:
\beq
 a_i &=& (h_i|_0H^{-1})\, \sigma  + {\cal O}(H^0), \\
 Z_i &=& Z_{i,-1}H^{-1} + {\cal O}(H^0)
 = 2s^{ijk}h_j|_0h_k|_0\,H^{-1} + {\cal O}(H^0), \nonumber \\
 \sigma &=& \sigma_0 + H \sigma_1 + {\cal O}(H^2), \nonumber \\
 k &=& (k_{0,-2}H^{-2} )\,\sigma  + (k_{0,-1}\sigma) H^{-1}\,
 + k_{0,0}\,\sigma + k_a|_0dx^a + {\cal O}(H). \nonumber
\eeq
For notational simplicity in section \ref{htozero} we will refer to all
non-singular variables (such as $k_a$ or $h_i$) by there values at
$H=0$ without including a subscript.

\subsection {The C-field and Metric as $H\to 0$}
\label{htozero}

A quick inspection of the C-field in eq.(\ref{cfield}) near the domain
wall shows that potential singularities appear only in the purely spatial part, and only at
order $H^{-1}$. The dangerous terms are of the form
\be
H^{-1}\left(2h_i - (Z_i^{-1} k_{0,-2}
\right)\sigma_0\wedge\textrm{Vol}_{T^2_i}
\ee
This singular term goes like (with no sum on $i$):
\be
2h_iH^{-1} - s^{ijk} H\(2h_j h_k + {\cal O}(H)\)^{-1}
H^{-2}\(4s^{ijk}h_ih_jh_k + {\cal O}(H)\) = 0\cdot H^{-1} + {\cal
O}(H),
\ee
so it cancels out generically.
The metric has singularities of leading order $H^{-2}$ and
subleading order $H^{-1}$, coming from the part of
the metric which contains $\sigma$
\be
- (Z_1Z_2Z_3)^{-2/3}\,k^2 + (Z_1Z_2Z_3)^{1/3} H^{-1} \sigma^2
= (Z_1Z_2Z_3)^{-2/3} \(-k^2 + (Z_1Z_2Z_3)H^{-1}\sigma^2\)
\ee
Up to a finite coefficient, the singular piece is proportional to:
\be
H^{-2}\(-k_{0,-2}^2 + (Z_1Z_2Z_3)|_{-3}\)(\sigma_0^2 +
2H\sigma_1\sigma_0) + H^{-1}\(-2k_{0,-2}k_{0,-1} +
(Z_1Z_2Z_3)|_{-2}\)\sigma_0^2
\ee
We can see that these singular terms vanish for our ansatz, since:
\beq
-(k_{0,-2})^2 + (Z_1Z_2Z_3)|_{-3}
&=& -64 h_1^2h_2^2h_3^2 + (4h_2h_3)(4h_1h_3)(4h_1h_2)=0, \\
-2k_{0,-2}k_{0,-1} + (Z_1Z_2Z_3)|_{-2} &=& - 16h_1h_2h_3(M_ih_i) +
 (M_1(4h_1h_3)(4h_1h_2) + \textrm{perms}) = 0. \nonumber
\eeq

\subsection{Zeroes of the $Z_i$ }

Looking at eq.(\ref{cfield}) or the determinant of the metric,
$\sqrt{-g_{11}} = (Z_1Z_2Z_3)^{1/3}\,H\,\sqrt{g_{\R^3}}, $ we see
that to avoid singularities it is necessary for $Z_i \neq 0$. We
have a simple tactic for enforcing the non-vanishing of the $Z_i$:
we demand that they obey
\be
\label{zbound}
Z_i H >0 \qquad \forall i \in {1,2,3}
\ee
everywhere.  We choose a convention where the $Z_i$'s are negative
at infinity, so this means that in Region I the $Z_i$ must remain
positive, while in Region II this means that the $Z_i$ must remain
negative. This uniformity also implies that $(Z_1Z_2Z_3)^{1/3}$ is
negative inside and positive outside, and guarantees that
combinations like $(Z_1Z_2Z_3)^{-2/3}$ and
$Z_i^{-1}(Z_1Z_2Z_3)^{1/3}$ remain positive everywhere. Using
eq.~(\ref{zform}), we see that the bound above can be written as:
\be
\label{fbound} 4f - s^{ijk}\sum_{p,q}  \frac{d^{pq}_j d_k^{pq}
\Pi^r_{pq}}{4n_pn_{q}} > 0
\qquad \forall i \in {1,2,3}.
\ee
In section~\ref{basic}, we will show that with two poles, this bound
is automatically satisfied modulo some relative sign conditions on
the $d^{pq}_i$. However, in general eq.(\ref{fbound}) will provide a
non-trivial constraint on the relative positions of the $\vec{x_p}$;
if solutions exist these conditions will define boundaries for the
moduli space of solutions.

\subsection{Closed timelike curves and horizons}

  Naively the $\sigma$ fibration has the potential to become timelike, thus
creating closed timelike curves.  To avoid this, we need to make sure the
negative term proportional to $k^2$ doesn't overwhelm the positive
term from the base space.  This means that we need to keep:
\be
-(Z_1Z_2Z_3)^{-2/3}\Big(k_0^2 - Z_1Z_2Z_3H^{-1})\Big) \ge 0,
\ee
so that the norm of the vector in the $\sigma$ direction remains
spacelike. Since the $Z_i$ have been tuned to avoid any
poles except at $H=0$,
and since the prefactor above is always positive, we only have to
worry about the second term in the product. In general this is a
complicated function, however, we can still exclude CTC's along the
$\sigma$ fiber in the neighborhood of the poles of $H$, $r_p=0$.
There it is easy to see using eq.(\ref{fixdistance}) that $k_0^2$ vanishes faster than $r_p$,
insuring that loops along the $\sigma$ fiber will remain space-like.

   We will exclude more general CTC's in our five dimensional
reduced space by requiring that this space be {\em stably causal},
i.e we will demand that there exists a smooth time function whose
gradient is everywhere timelike \cite{hawkingellis}. Our candidate
function is the coordinate $t$, and it qualifies as a time function
if
\be
\label{tnorm}
-g^{\mu \nu} \partial_\mu t \partial_\nu t=-g^{tt} = \(Z_1Z_2Z_3\)^{-1/3}H^{-1}\((Z_1Z_2Z_3)H - H^{2}\,k_0^2 -
g^{ab}_{\R^3}k_ak_b\)
>0
\ee
We will leave for future work the question of possible extra
constraints on the relative pole positions which come from this
condition.

Granted the time function $t$, we can now proceed to show that there
are no event horizons.  The vector $\partial_r$ has a norm,
\be
g_{rr} = (Z_1Z_2Z_3)^{-2/3}\((Z_1Z_2Z_3)H - k_r^2\) \ge -g^{tt},
\ee
which is positive everywhere due to eq.(\ref{tnorm}).  Consider the
following vector field:
\be
\xi = \( {g_{rr}\over - g^{tt}} \)^{1/2} g^{t\mu}\partial_\mu +
    \epsilon\, \partial_r.
\ee
The norm of $\xi$ is
\be
\|\xi\| = - g_{rr}\(1 - \epsilon^2\)
\ee
For $\epsilon < 1$ this is always negative, therefore trajectories generated
by this vector field  will always be timelike.  If we also choose $\epsilon > 0$, these
trajectories will always eventually reach asymptotic infinity and so
there can be no event horizon.

\subsection{Topology of the $\sigma$-fibration}

The $\sigma$-fibration is preserved after the corrections in the
full metric. Notice that the base metric has orbifold points with
identification on the $\sigma$ fiber.  The order of these points can
be determined by the following calculation.  For any given
two-sphere on the base $\R^3$, we can determine the first Chern
class, $c_1$, of the $\sigma$ fibration $U(1)$ bundle by integrating
$d\sigma$ over that 2-sphere and then using Stoke's theorem to turn
that into an integral over the inside $B^3$:
\beq
\int_{S^2} d\sigma  &=& \int_{S^2} \star _3 dH
= \int_{B^3} d\star _3dH = \sum_p \int_{B^3} n_p
\,\delta^3(\vec{r}-\vec{r_p}).
\eeq
This yields an integer which counts the poles inside of $S^2$.  If this
integer is zero, the topology of the $\sigma$-fiber over this $S^2$ is
$S^2\times S^1$, if the integer is
$\pm1$ then the topology is that of $S^3$. Any larger integer $m$ will
give the topology $S^3/Z_m$.

If we want to understand the corrections to the fibration from the
whole metric, we rewrite the $\sigma$-fiber piece as:
\be
(Z_1Z_2Z_3)^{-2/3}(Z_1Z_2Z_3H^{-1} - k_0^2)(\sigma
 - Ak_adx^a)^2, \qquad
A = \frac{H k_0}{Z_1Z_2Z_3 - H k_0^2}.
\ee
For a given $S^2$, the correction to $c_1$ of the $\sigma$ bundle is
\beq
\int_{S^2} d(Ak_adx^a) &=& \int_{S^2} (dA\wedge k_adx^a
+ A \wedge d(k_adx^a)) \\
&=& \int_{B^3} (d^2A) \wedge k_adx^a + A \wedge d\star _3(H\,
d\bar{K} - \bar{K}\,dH + h_i\,dZ_i - Z_i\,dh_i) = 0. \nonumber
\eeq
The first of these terms is trivially zero, while the second
vanishes due to eq.~(\ref{Kpoles}).  Therefore, there are no
corrections to the $\sigma$-fiber's topology.

\subsection{Topology of the Gauge Fields}

Another interesting topological aspect for our solution is the topology of
the C-field.  We can gain a clearer picture of this by considering a membrane, labelled
$\M_i$ wrapped on the torus $T_i$.  This effectively yields a charged particle in the
five-dimensional reduced space with charge,
\be
e_i = V_i\, \tau_2 = \frac{V_i}{(2\pi)^2 \ell_P^3},
\ee
which experiences a gauge-field and field strength:
\be
\A_i = 2a_i - Z_i^{-1}(dt + k),\qquad F_i = d\A_i.
\ee
For quantum consistency of the wave function for probe charges
$e_i$, we usually require that on the five-dimensional space that
the field strength be an integral cohomology class; the properly
normalized integral of this class on a regular two-cycle should
yield an integer.  The compact two-cycles in our geometry, $S^{pq}$,
are represented by line segments on $\R^3$ between two points
$\vec{x}_p$ and $\vec{x}_q$ where the function $H$ blows up, along
with the fiber $\sigma$. Of course, if either $n_p$ or $n_{q}$ is
larger than one, the corresponding $S^2$ will have orbifold
singularities. This means we should look for an integral cohomology
class on the universal cover of $S^{pq}$, i.e. $n_p\cdot n_{q}$
times our original cycle.

  The arguments above lead us to define an integer for each two-cycle,
$S^{pq}$, derived from the following flux integral (all forms are
pulled back to the two-cycle):
\be
m^{(pq)}_i = n_p\,n_{q}\, {e_i\over 2\pi}
\int_{\vec{x}_p}^{\vec{x}_q} \int_\sigma \, F_i d\tau ds = 2
n_p\,n_{q}\, e_i \;\A_i\Big|_p^{q} = 2 n_p\,n_{q}\, e_i\,(2a_i -
Z_i^{-1}k_0)\Big|_p^{q}
\ee
To evaluate this, we can use the fact that $k_0 \to 0$ at the points
where the $\sigma$ fiber degenerates.  Thus,
\be
\label{H2} m^{(pq)}_i = n_p\,n_{q}\, e_i \,\Big({d^{\cdot}_i\over
n_p}\Big)\Big|_p^{q} = e_i\Big(n_{p} \,d^{(q)}_i - n_{q} \,d^\p_i
\Big) = e_i d_i^{pq}.
\ee
Our story, however, does not end here.  Near each orbifold point
$p$, as mentioned above, the local geometry is a cone over
$S^3/\Z_{n_p}$. This has $\pi_1 = Z_{n_p}$  and implies that we have
the possibility of a discrete Wilson line for each gauge field with
a phase of the form $2\pi\, m^\p_i/n_p$ where $m^\p_i \in \Z$
(essentially we are using the fact that $H^2(S^3/\Z_{n_p}) =
Z_{n_p}$). The invariance of the local gauge field under a shift of $m_i^{(p)}$ by $n_p$ can be implemented by a gauge transformation similar to that of eq.(\ref{gauge}). Looking at the gauge field $\A_i$ near one of the
orbifold points, for example along the $\sigma$ fiber, we see that
the Wilson line phase is:
\be
 2\pi\,\( 4 e_i h_i /H \)\Big|_p =
 2\pi\, m^\p_i/n_p.
\ee
This implies a quantization for the $d_i^\p$ of the form:
\be
d^{(p)}_i = m^{(p)}_i/e_i = \frac{(2\pi)^2 \ell_P^3}{V_i}\,
m^{(p)}_i,
\ee
with several consequences.  First, the cohomology requirement in
eq.(\ref{H2}) is trivially satisfied: $m_i^\pp = n_pm^{(p')}_i -
n_{p'} m^\p_i$.  Second, unless all the $m^\p_i$ at a given point
$p$ are multiples of $n_p$ or on one of it's divisors, the
singularity is ``frozen'' or partially ``frozen'' \footnote{One can
see this by reducing the M-theory solution down to IIA on the
appropriate circle, then the Wilson line gives mass to otherwise
twisted closed strings, or by dualizing to IIB via the relevant
$T^i$'s and then the dual circle will be non-trivially fibered so
that there is no longer an orbifold point. For similar ideas see
\cite{fluxes,fluxes2}}. Finally, if we use our formula (\ref{fixq})
for the monopole charges, we get:
\be
Q^{(p)}_i = -{4G_5 e_i \over \pi} \, s^{ijk} \frac{m^{(p)}_j
m^{(p)}_k}{2n_p}.
\ee
and so the quantized membrane charge at infinity will be
\be
\label{nsum}
N_i  = \frac{\pi}{4 e_i G_5 } Q_i
    = \frac{\pi}{4e_iG_5} \sum_{p=1}^N Q^{(p)}_i
    = -\sum_{p=1}^N s^{ijk}\frac{m^{(p)}_j m^{(p)}_k}{2n_p}
    = -\sum_{p,q} s^{ijk}\frac{m^{pq}_j m^{pq}_k}{4n_pn_q}.
\ee
Note that the fact that the $N_i$'s are integers written as a sum of
rational numbers add further 
constraints on the $m_i^\p$'s or $m_i^{pq}$'s.

\subsection{Summary of Conditions}

  We finish this section by summarizing the exact conditions which
will define a smooth (modulo 
orbifold points) and regular
11-dimensional supergravity solution with three membrane charges and
4 supersymmetries, with no CTC's or event horizons.  The solution is
completely parameterized by a set of poles on $\R^3$ with quantized
residues $n_p$ and quantized fluxes $m^{pq}_i$.  These, and the
quantities that depend on them, must satisfy the following
conditions:
\beq
\label{c1}
  &1)& \sum_i d^\p_i \; + \sum_{q} {1\over 4 n_p^2
   n_q^2}{1\over r_{pq}} \prod_i d_i^{pq} = 0, \\
\label{c2}
  &2)& Z_iH>0
  \qquad \forall i \in {1,2,3}, \\
\label{c3}
  &3)& (Z_1Z_2Z_3)H - H^{2}\,k_0^2 - g^{ab}_{\R^3}k_ak_b > 0.
\eeq
These need not be independent conditions, for example it may be possible
that condition $(1)$ implies $(3)$ implies $(2)$.

 There exists a {\em canonical solution} for eq.(\ref{c1}), as long
as the $d_i^{pq}$ are such that the $r_{pq}$ come out non-negative, of the form:
\be
r_{pq} = {1\over 4 n_p^2n_q^2} {\prod_i d_i^{pq}\over \sum_i d_i^{pq}}.
\ee
It is not yet clear what extra conditions on the $d_i^{pq}$, if any,
are required for this canonical solution to satisfy the other two
constraint equations.

\section{The Basic 2-Pole Example}
\label{basic}

\begin{figure}[htp]
\centering \psfig{figure=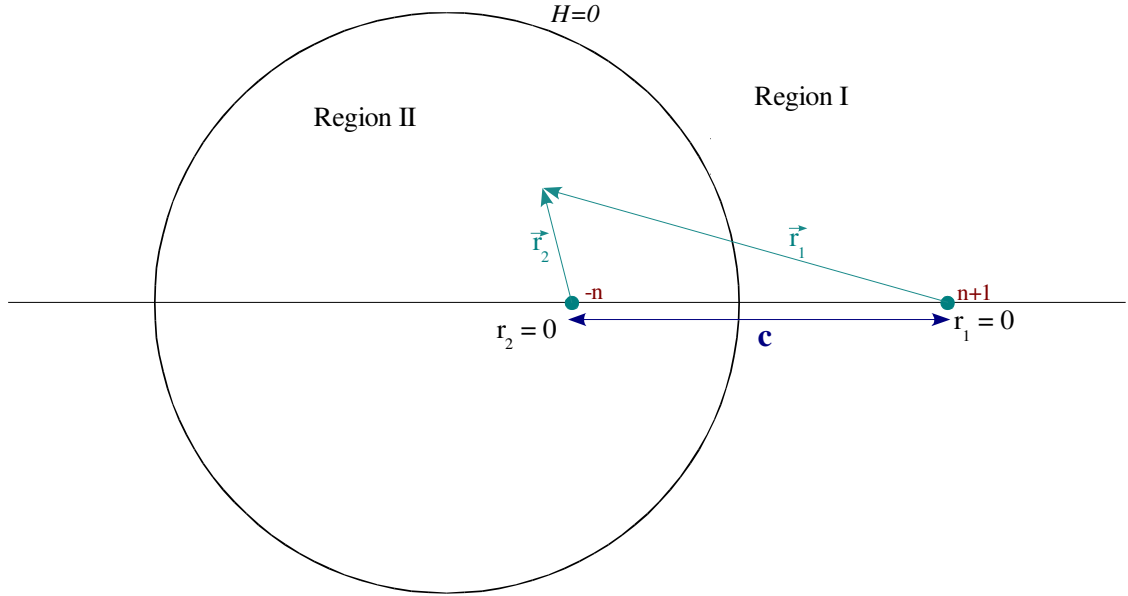,width=0.9\textwidth} \caption{Basic
2-pole example has a spherical domain wall}
\end{figure}

Now that we have worked out a general formalism, let us illustrate
it with the simplest example, the two pole solution;  this reproduces
the solutions of \cite{mathur1,mathur2,mathur3}.  The simplest
possible $H$ that interests us has two poles and can
be written in the following form (w.l.o.g we choose $n>0$):
\be
H = \frac{n+1}{r_1} + \frac{-n}{r_2}
\ee
where $r_2$ is defined as the distance in $\R^3$ from a point a
distance $r_{12}=c$ below the origin on the $\phi$ symmetry axis;
this gives $H\!K$ a $U(1)\times U(1)$ isometry.  Note that with this
choice of the function $H$, Region II is a spherical region with its
center just below the point $r_2 = 0$ (see figure).  The function
$f$ is $f = -n\,r_1 + (n+1)\,r_2$ and has as its minimum value
$-n\,c$.

For the dipoles we choose $d_i^{(1)} = - d_i^{(2)} = d_i$.  The
$Z_i$'s and the net charges at infinity now take the simple form:
\be
\label{charges2}
Z_i = 1 + {Q_i\over 4f},\qquad Q_i = {s^{ijk}\over
2}\,\frac{d_jd_k}{n(n+1)}.
\ee
Notice that to guarantee positive $Q_i$'s, we must impose that the
$d_i$'s all have the same sign.  From eq.(\ref{Kpoles}), we see
that the poles of $K$ at $r_1=0$ and $r_2=0$ are:
\be
\ell_1 = \frac{d_1d_2d_3}{16(n+1)^2},\qquad \ell_2 = -\frac{d_1d_2d_3}{16n^2}.
\ee
This gives a simple form to $k_0$:
\be
k_0 = \sum_id_i \frac{r_2 - r_1}{4f}
- \frac{\Pi_id_i}{n^2(n+1)^2}\frac{(n+1)r_2 + nr_1}{16f^2}.
\ee

\subsection{Solving the Constraints}

In this simple case, there is a unique solution to the first
constraints, eq.(\ref{c1}), which is the canonical solution:
\be
\label{c}
c = \frac1{n^2(n+1)^2} \frac{\Pi_i d_i}{4\sum_id_i} =  {1\over
4n(n+1)}\frac{Q_1Q_2Q_3}{(Q_1Q_2 + Q_2Q_3 + Q_3Q_1)}.
\ee
Using this value and the fact that the minimum value for $f$ is
$-n\,c$, the second constraint, eq.(\ref{c2}) can be easily checked.
The final constraint, eq.(\ref{c3}), after some algebra, can be put
in the form:
\be
\Big[ (M+
4f) + D/(2r_1 + 2r_2 + 2c)\Big]/4 r_1 r_2 >0
\ee
where $M = Q_1 + Q_2 + Q_3$ and $D = Q_1Q_2 + Q_2Q_3 + Q_3Q_1$. Each
term in this product is manifestly positive, so we are guaranteed a
smooth solution without CTC's or event horizons.

\subsection{Features of the Solution}

The angular momenta can be read off from eq.~(\ref{J}):
\beq
\label{JR}
J_R &=& - {d_1d_2d_3}\frac{2n+1}{n^2(n+1)^2}=
\pm(2n+1)\sqrt{Q_1Q_2Q_3\over n (n+1)} \\
\label{JL}
J_L &=& 4c\sum_id_i = \pm 4c \sqrt{n(n+1)\over Q_1Q_2Q_3}(Q_1Q_2 +
    Q_2Q_3 + Q_1Q_3) = \pm \sqrt{Q_1Q_2Q_3\over n (n+1)}.
\eeq
The whole of the solution is completely determined by an integer $n$
and the three $d_i$'s or, more physically: $n$, the three charges
$Q_i$ and a $\pm$ sign derived from a collective sign choice for
$d_i$'s.  We can extend this solution to negative $n$'s by taking $n
\to - (n+1)$, which just flips the sign of $J_R$.

In general, if all three charges $Q_i$ are of order $Q$, large
relative to $\ell_P^2$, then our system will be macroscopic for
finite $n$, since $c \to Q/n(n+1) >> \ell_P^2$.  If one of the
charges is much smaller than the others, e.g. $Q_3 << Q_1,Q_2$, then
$c$ will be of order that smaller charge, and there will be high
curvature regions if we don't also have $Q_3 >> \ell_P$.

To get a further understanding for the features of the two-pole
solution, we draw the readers attention back to last part of
eq.(\ref{charges2}), which can be rewritten as
\be
N_i = {s^{ijk}\over 2} {m^{12}_jm^{12}_k\over n(n+1)}
    = {s^{ijk}\over 2} {m_jm_k\over n(n+1)}.
\ee
Since the $N_i$ are integers, the product of any two integer $m_j$'s
must be divisible by $n(n+1)$. Define $a = gcd\Big(m_3,n(n+1)\Big)$
and it's complement $b= n(n+1)/a$.  The divisibility condition on
the $m_i$'s implies that there exist integers $k_i$ and $k$ such
that
\be
m_1 = b\,k_1,\; m_2 = b\, k_2,\; m_3 = a\, k_3,\;\; \textrm{and}\;\;
k_1\,k_2 = a\, k.
\ee
These yield the following relations
\beq
\label{Ns}
 && N_1 = k_2\,k_3,\; N_2 = k_1\,k_3,\; N_3 = b\,k,\;  N_1 N_2 =
(k\,k_3)\,m_3, \\ \nonumber
 && j_L = {\pi\over 4G_5} J_L =(k\,k_3),\qquad j_R = {\pi\over 4G_5} J_R = (2n+1)(k\,k_3).
\eeq
Clearly the quantity $kk_3$, let us label it $N_s$, plays a
prominent role.

If $k_3 =1$ and $a$ divides $n$, there is a natural way of thinking
about the numbers above.  We take the third torus $T_3$ to have very
small volume $V_3$. We then get a dual IIB description with $N_1$
and $N_2$ three-branes intersecting over a circle of radius
$\ell_P^3/V_3$ with $N_3$ units of momentum going around the circle.
They can intersect over one long string, wrapped $N_1 N_2$ times
around that circle or, more generally, $N_s = N_1 N_2/m_3$
substrings wrapped $m_3$ times.  If we start with no momentum, we
have:
\be
j_L = j_R = N_s = N_1N_2/m_3 = kk_3
\ee
Following steps summarized in \cite{mathur2} we can apply a spectral
flow shift of $2\tilde{n}$ to this CFT and get shifted values:
\be
j_R \to j_R + 2\tilde{n}\, N_1 N_2,\qquad N_3 \to N_s\,
\tilde{n}\,(\tilde{n}\,m_3 + 1)
\ee
Now define $n = \tilde{n}\,m_3$ and we get:
\be
N_3 = n(n+1) N_s/m_3,\qquad j_L = N_s, \qquad j_R = (2n+1) N_s,
\ee
which fits nicely with eq.~(\ref{Ns}), except with the stronger
assumption that $m_3$ divides $n$.   Thus the states \cite{mathur2}
are in a subclass of the generic two pole solution.

\subsection{The Two Pole Solution as a Microstate}

  Since our original motivation was to find microstates for black
holes and black rings, it is natural to ask if this geometry has the
same asymptotics as one of those black objects.

Let us first consider the BMPV black hole~\cite{BMPV}.  Immediately,
we run into a a problem: the basic solution above always has a
non-zero $j_L$, so strictly speaking cannot be considered a
microstate for the BMPV black hole in~\cite{BMPV}, which has $J_L
=0$.  For large values of $n$ and small values of $N_s$, however, it
is possible to have $J_R$ be macroscopically large in Planck units,
while $J_L$ remains of order the Planck length and can be
coarse-grained away. In fact, the states that dominate the partition
function for the CFT dual of the black hole (see \cite{juanthesis}
for a review) are in the single ``long string'' phase, which has
$j_L = +1$.  Taking this caveat into account, we see that in the
large $n$ limit $J_R^2 \to 4Q_1Q_2Q_3$, which means that our two
-pole states are at best microstates of a zero area BMPV blackhole
and as such cannot be considered as microstates for a proper BPS
blackhole. We must look to other black objects.

  By construction, the two-pole solution has $\vec{D}_i$'s which are
parallel.  This feature potentially qualifies it as a black ring
microstate.  Black rings, however, have more parameters than black
holes, so we have to be careful in matching these.  If we use the
nomenclature in~\cite{matring2}, we see that the parameters $q_i$
which appear in the expression for the horizon area only appear as
$R^2q_i$ in the asymptotics, where $R$ is the inner radius of the
ring. This leads to an ambiguity, since we match
\be
\label{matchc}
 4c d_i = R^2 q_i.
\ee
As it turns out, $q_i$ and $d_i$ are quantized in the same units, so
we have
\be
\label{ratio}
 R^2/4c = d_i/q_i = a/b,\qquad  a,b \in \Z.
\ee
The black ring solutions require
\be
\label{bound} \(Q_i = \frac{s^{ijk}}{2}\frac{d_jd_k}{n(n+1)}\) \ge
\frac{s^{ijk}}{2}\,q_jq_k
 \;\; \Rightarrow\;\; \frac{a^2}{b^2} \ge
n(n+1).
\ee
Without going into too much detail, if we compare our expressions
for the angular momenta eq.(\ref{JR}) and eq.(\ref{JL}) with those
in \cite{matring2,BW}, we find that the expressions for $J_L$ match
(this match is independent of $N$) if we use eq.(\ref{matchc}) but
that our $J_R$ is always too small to qualify for a black ring. Only
in the limit where we saturate the bound in eq.(\ref{bound}) and $n$
is very large do we start converging on the same value for $J_R$. At
that point, our solution ends up sharing the asymptotics of a
zero-area ring.  This is a very similar story to that of the BMPV
black hole.

  In conclusion, the two-pole solution can at best appear as a
microstate for 
``black" rings and ``black'' holes with
zero horizon areas.  In the context of the CFT microstates
connection we developed earlier, this is none too surprising, as
those states are unique given their charges and angular momenta. We
also see that our supergravity solution does not appear to have any
adjustable parameters, hence should not belong to the very numerous,
i.e. entropic, class of states that should match a black object.

\subsection{The Domain Wall}

The surface $H=0$ has a radius:
\be
  R^2_{ DW} = 4r_{DW} = 4c \,{Q_1Q_2Q_3\over J_LJ_R}.
\ee
In the region near $r_2=0$ with radius of order $R=\sqrt{c}$, the
functions $Z_i$ are locally constant and $k \propto r_2\, \sigma$,
which means that the local metric is a $Z_n$ orbifold of a G\"{o}del
universe. This suggests that our solution is in fact the un-smeared
version of the hypertube speculated about in \cite{GH} and is in
fact a smooth resolution of the type of domain wall illustrated
there.

  Given that the hypertube-like solution we have found has completely smooth
supergravity fields except for mild orbifold singularities, one is
tempted to ask what physical features the surface at $H=0$
exhibits, if at all, to mark it's presence.  An associated issue is
the question of just where the membrane charge detectable at
infinity is sourced.

To answer the first question, it is interesting to consider a probe
membrane in our background wrapping the $z_1,z_2$ and $t$
directions.  If we include a small velocity $v^0$ along the $\sigma$
fiber and $\vec{v}$ in the base space, we see that the probe action
in the static gauge has an expansion which looks like:
\beq
&& e_1\int  \Bigg[Z_1^{-1}\sqrt{(1 + k_0v^0 + k_av^a)^2 -
Z_1Z_2Z_3(H^{-1}v_0^2 + H g_{ab}v^av^b)} \\
&&- Z_1^{-1}(1 + k_0v^0 + k_av^a) + 2a_{i0}v^0 +
2a_{ia}v^a\Bigg]d\lambda.  \nonumber
\eeq
Clearly, there is no potential term, as can be expected for a BPS
configuration.  Away from the $H=0$ domain wall, we expand the
action in small powers of the velocity, and we get:
\be
e_1\int \Big[-{1\over 2}\,{Z_2Z_3}\( H^{-1}v_0^2 + Hg_{ab}v^av^b \)
- 2(a_{i0}v^0 + a_{ia}v^a) + {\cal O}(v^3)\Big] d\lambda.
\ee
This is the action for a charged particle on $H\!K$ traveling in a
magnetic field, albeit with variable mass.  Near $H=0$,
the mass parameter and magnetic field blow up! At this point,
we can no longer work in the small $v$ approximation and
must go back to the full action, itself always finite.

   Another way to analyze the action as  $H \to 0$, is to realize that
$\partial_t$ becomes lightlike turning our static gauge into a light
cone gauge. Expanding the worldline tangent for our wrapped membrane
``particle'' about the lightlike velocity vector $\partial_t$ gives
us a theory with exact Galilean invariance. The momentum $p_t$
becomes the lightcone energy, and since the only mixed term in the
line element looks like $dt\,\sigma$, $\tau$ seems to be the natural
lightcone time variable.  The interesting thing here is that $\tau$
is periodic, and $\partial_{\tau}$ is spacelike.  Thus our membrane
worldtheory takes on aspects of a finite temperature
non-relativistic system!  It is interesting to speculate about what
this might mean for the process of ``heating up" our solution to
make it a microstate of a finite temperature near-BPS black hole,
something like the states considered in \cite{rossnonsusy}.  We
leave this for future work.

  In summary, the domain wall at $H=0$ leads to some interesting
behavior for the worldvolume on our probe brane, yet there is still no
real discontinuity there and no vanishing of the probe kinetic term
as for an enhancon~\cite{JPP}, in fact just the opposite.

  The charge in the hypertube has dissolved away, so where should we think of it
existing?  In our case, an alternate scenario to charges localized on
a domain wall appears.  We see instead a situation similar to that of the
geometric transition in \cite{GV,Vafa}, where a large number of D-branes
wrapped on an $S^3$ at the tip of a cone over $S^2$ is replaced by
flux on a non-contractible $S^2$ at the tip of a cone over $S^3$.
We can take a decoupling limit for our solution by removing the $1$'s in
the $Z_i$'s to get a five-dimensional space which is asymptotically an
orbifold of $AdS_2\times S^3$.  The infrared limit of the
holographically dual theory should reflect the appearance of this
geometric transition.

\section{Adding More Poles}
\label{more}

    Clearly, if we want to achieve a more general solution which
could have the same asymptotics as BPS black hole or black ring with
a non-zero horizon area, the next step is to consider a solution
with more poles, starting with three poles. We will see that this
type of solution has more adjustable parameters, making it a
more likely
candidate.

\subsection{Simple Three Pole Solutions}

    With three poles to play with, the simplest scenario is for $H$
to have two poles with residue $n_1=n_2=+1$ and one with residue
$n_3 = -1$. With these choices, the solution is completely smooth!
We define the three radial distances from these poles as $r_1,r_2$
and $r_3$, and corresponding dipoles $d^{(1)}_i,d^{(2)}_i$ and
$-(d^{(1)}_i + d^{(2)}_i)$. One nice feature is that we can now vary
the dipoles and $J_L$ by our choice of pole arrangements

We can further simplify our solution by working with dipoles that
are diagonal in the three $U(1)$'s.  This will guarantee that the
dipole vector's $\vec{D}_i$ will all be parallel, and thus allow
comparison with BPS black rings, especially the ones in
\cite{matring1}.

We start with probably the simplest example. $d_i^{(1)}= d_i ^{(2)}
=d$ and $d_i ^{(3)} = -2d$. Then we find the set of equations
eq.~(\ref{c1}) become simply
\beq
r_{13}=r_{23}=\frac{d^2}{12} ,
\eeq
with no restrictions on $r_{12}$. We must now satisfy $Z_i H>0$,
eq.~(\ref{c2}). This condition becomes
\beq
4(r_2 r_3 +r_1 r_3 -r_1 r_2) + d^2 (r_1 +r_2) > 0 .
\eeq
Using our values for $r_{13}$ and $r_{23}$ one can easily show this
equation is satisfied at the three poles. It is also trivially
satisfied at asymptotic infinity. Though it is difficult to show
analytically, a numerical analysis confirms that it is satisfied
everywhere for any value of $r_{12}$. Let us examine the
implications of this. Without loss of generality we choose to place
pole 3 at the origin, and pole 1 on the z-axis. Then, since there
are no restrictions on $r_{12}$ we are free to place it anywhere on
a circle of radius $r_{23}=r_{13}=d^2/12$. Let the angle 132 between
segments ${13}$ and ${23}$\  be called $\psi$. Then we find
\beq
\vec{D}_i = \frac{d^3}{12} [ \hat{z} (1+\cos\psi)+\hat{x} \sin \psi]
.
\eeq
One can read off the asymptotic charges for this solution. They are
\beq
Q_1=Q_2=Q_3=2d^2 , \ \ J_R^2=36 d^6 , \ \ J_L^2 =2d^6 (1 + \cos
\psi) .
\eeq
One can easily see that while $Q_i$ and $J_R$ are independent of our
pole arrangement, $J_L$ is very sensitive to it. We find that $0
\leq J_L^2 \leq 4 d^6$, where the minimum value is reached when the
negative pole (3) is in the middle of the two positive poles. The
maximum value is reached when the two positive poles lie on top of
each other, returning us to the 2-pole example from the previous
section with $n=1$.  Note also that even in this very simple three
pole case, the last condition, eq.~(\ref{c3}) is quite challenging;
and we will not check it here.

    For the purposes of finding microstates for black objects,
adding a third pole already seems to help.  For example, we can
explicitly set $J_L$ to zero, which is characteristic of BMPV
microstates.  Unfortunately $J_R$ is still too large and saturates
the BMPV bounds just as in the two-pole case.  On the black ring
side, things look more promising.  As we adjust the angle $\psi$ the
three dipoles have variable magnitude $2d\cos\psi/2 \leq 2$,
matching to a class of black rings with variable $J_R$, a function
of the dipoles and smaller as we increase $\psi$.  The upshot is
that for $0 < \psi < \pi$, we can now adjust the constant $a/b$ in
eq.(\ref{ratio}) so that the three-pole solution's quantum number
$j_R$ matches that of the black ring with the same dipoles and
charges at infinity. Hence, modulo the CTC condition in
eq.(\ref{c3}), we have identified supergravity microstates for black
rings with finite-sized horizons.

\subsection{Comments on the General Case}

  As we have seen, the three pole case allows us
more freedom in positioning our poles.  This freedom tends to only
vary $J_L$ and the $\vec{D}_i$'s, hence it does not generate a large
class of states with the same asymptotics.  Ideally, we would like
to be able to fix the $Q_i$'s, $\vec{D}_i$'s, $J_L$ and $J_R$ and
still find many solutions. It seems reasonable to believe that
adding more poles should allow us to have much larger moduli spaces
of solutions, including substantial subspaces with the same
$Q_i$'s,$D_i's$ and $J_{L,R}$; further work towards understanding
BPS black ring and black hole microstates will requiring developing
a better understanding of the general $n$-pole solution.

\section{Discussion}

    The general picture that is emerging from our solution is
that of generic five dimensional BPS three-charge black hole and
black ring microstates coming from a harmonic function with a large
number of poles.  These solution are characterized by discrete
choices, the $d_i^{pq}$'s, as well as continuous ones, the relative
positions of the points $\vec{x_p}$.  This dichotomy is reflected in
the angular momenta, with $J_R$ clearly discrete in all cases while
$J_L$ varies continuously requiring explicit quantization.
Similarly, we see one set of exact constraints eq.(\ref{c1})
complemented by inequalities, inexact constraints, from
eqs.(\ref{c2}-\ref{c3}).

    It is likely that the dichotomy above arises from the explicit $U(1)$
invariance that permeates our solution, arising from our ansatz for
the pseudo-hyperkahler base space in the Gibbons-Hawking form. This
raises the question of what more general pseudo-hyperkahler base
spaces look like.  These general spaces could also have integer
quantized fluxes on two-cycles, but these need no longer share the
same $U(1)$ symmetry and so the asymptotic angular momenta would
both vary continuously.

  The exact $U(1)$ symmetry which we have, non-generic in five
dimensions, can become an asset if we choose to use it to reduce to
four dimensions.  This can be done by simply adding a term of the
form $1/R^2$ to the harmonic function, and relaxing the condition on
the sum of the residues; most of our analysis will survive
unchanged. This yields a substantial generalization of the four
dimensional solutions in \cite{GSY, EEMR, BKW}. In those papers,
solutions appear with a KK-monopole charge which is basically
$\sum_p n_p$, but only allowing positive residues!  Modulo
constraints of the form in eqs.(\ref{c1}-\ref{c3}), the possibility
of adding negative residues to the poles of $H$ greatly broadens
this class of supersymmetric four dimensional solutions.  If we make
$\tau$ the M-direction, the appearance of negative poles corresponds
to having anti-D6-branes while keeping the supersymmetries of
D6-branes, this is similar to what happens in~\cite{DDbar}. The
appearance of both D-branes and anti-D-branes in a black hole
microstate is something we have learned to expect in near-extremal
systems to but is new to extremal ones.

   We close by stressing what we feel is the most important idea
emerging from our analysis: supergravity solutions dual to wrapped
brane bound states replace a core region which naively would have a
naked singularity or a horizon with a core region containing a
``foam" of new topologically nontrivial cycles. This is a
fascinating combination of Vafa etal.'s geometric transitions
picture~\cite{GV,Vafa} and melting crystal space-time foam
picture~\cite{crystal,foam} which will certainly attract future
attention.

\section*{Acknowledgements}
E.G. would like to dedicate this paper to Jean-Paul Gimon, loving
father and friend.

We would like to thank V.~Balasubramanian, I.~Bena, D. Berenstein,
P.~Ho\c{r}ava, D.~Mateos, E.~Sharpe, J.~Simon and N.~Warner for
useful conversations, as well as the organizers of the ``QCD and
String Theory" workshop at the KITP. PB would like to thank the
organizers of the String Phenomenology workshop at the Perimeter
Institute for a stimulating environment where some of this work was
done. PB is supported by NSF grant PHY-0355074 and by funds from the
College of Engineering and Physical Sciences  at the University of
New Hampshire. EG is supported by the Department of Energy under DOE
contract numbers DE-FG02-90ER40542 and DE-AC03-76SF00098. TSL would
like to thank the Kavli Institute of Theoretical Physics and the
Graduate Fellows program for warm hospitality and a stimulating
environment during the early stages of this work. TSL was supported
in part by the KITP under National Science Foundation grant
PHY99-07949, the National Science Foundation under grants
PHY-0331728 and OISE-0443607, and the Department of Energy under
grant DE-FG02-95ER40893.



\bibliography{DW}
\bibliographystyle{utphys}

\end{document}